\documentstyle[preprint,prb,aps]{revtex}
\begin{document} \draft
\title{
Topological Defects, Orientational Order, and Depinning of the
Electron Solid in a Random Potential
}
\author{Min-Chul~Cha and H.A.~Fertig}
\address{
Center for Computational Sciences and
Department of Physics and Astronomy, University of Kentucky,
Lexington, Kentucky 40506-0055.}
\address{\rm (Draft \today)}
\maketitle
\begin{abstract}
We report on the results of molecular dynamics simulation (MD) studies
of the classical two-dimensional electron crystal in the
presence disorder.  Our study is motivated by recent experiments
on this system in modulation doped semiconductor systems in
very strong magnetic fields, where the magnetic length is much
smaller than the average interelectron spacing $a_0$, as well
as by recent studies of electrons on the surface of helium.
We investigate the low temperature state of this system using
a simulated annealing method.
We find that the low temperature state of the system always
has isolated dislocations, even at the weakest disorder
levels investigated.
We also find evidence for a
transition from a hexatic glass to an isotropic glass as
the disorder is increased.  The former is characterized
by quasi-long range orientational order, and the absence
of disclination defects in the low temperature state,
and the latter by short range orientational order and
the presence of these defects.  The threshold electric
field is also studied as a function of the disorder strength,
and is shown to have a characteristic signature of the
transition.  Finally, the qualitative behavior of the electron
flow in the depinned state is shown to change continuously
from an elastic flow to a channel-like, plastic flow as the
disorder strength is increased.
\end{abstract}

\narrowtext
\section{Introduction}
\label{sec:introduction}

Sixty years ago, Wigner\cite{wigner} predicted that at low enough densities,
the ground state of a collection of electrons should have a crystalline
form.  Convincing evidence of this Wigner crystal (WC) was first reported
by Grimes and Adams\cite{grimes} forty-five years later for a collection
of electrons on the surface of He$^4$.  This system is at such
low densities that exchange effects among the electrons are
negligible, so that for all intents and purposes, it
may be regarded as purely classical.

More recently, there have been a number of studies searching for the
WC in semiconductor systems, where considerably higher densities may
be obtained.
A major drawback of semiconductors is that one usually needs
to introduce (randomly located) dopants in order to put free electrons into
the system.  These dopants then become a source of disorder,
which competes with the intrinsic tendency of low-density
electrons to form an ordered solid in their ground state.
Indeed, in a bulk doped semiconductor, electrons usually
bind to the charged donors at zero temperature, forming a
completely {\it disordered} state in this limit.

Great progress has been made on semiconductor systems
since the invention of modulation-doping\cite{stormer}.
In this class of materials, a two-dimensional layer of electrons
(``two-dimensional electron gas'', or 2DEG) is
collected at the interface of two different semiconductors
(typically, GaAs and AlGaAs), with the donor atoms setback
by some distance $d$ from the electron layer.
By fabricating a sample with a large ratio $d/a_0$,
where $a_0$ is the lattice constant of a perfect electron lattice,
the greatest source of randomness is relatively
far from the electrons, and disorder effects become much less
pronounced.  (Indeed, in situations where they do not
crystallize,
the electrons have remarkably high mobilities,
suggesting that the electrons undergo very little elastic
scattering\cite{goldman,tsui}.)  This sytem
has thus become one of the leading candidates for the
observation  of a WC.

This paper studies, using computer simulations, the low
temperature properties of the two-dimensional WC, in the low
density limit, where the electrons may be treated as
classical.  Many of our results will also be applicable
for higher density electron systems in the presence of
very strong magnetic fields.  Our primary focus
will be on low temperature configurations and how they
are affected by disorder, and on the behavior of
the electrons in the presence of an electric field
(i.e., depinning properties.)

The application of a strong, perpendicular magnetic field
further enhances the prospects of stabilizing a WC in the 2DEG.
In the absence of a magnetic field, the
zero-point kinetic energy cost
of localizing electrons at lattice sites competes with
the potential energy gained by forming a lattice; at high
enough densities, the former will always destabilize
the lattice.  By contrast, in a magnetic field $B$ electrons may be
localized to within a length scale $\l_0= \Bigl( {{\hbar c} \over
{eB}}\Bigr)^{1/2}$,
and simultaneously have their lowest possible kinetic
energy.
Since $l_0$ becomes arbitrarily small in the high field limit,
one may form highly localized wavepackets, from which a WC
state may be constructed.  Furthermore, for $\l_0 \ll a_0$, exchange
effects are negligible\cite{maki},  and the ground state of the
system may in a sense be regarded as the {\it classical} ground state
of the electron system: the electron wavepackets form a
crystalline state, and the potential energy of the configuration
is given to an excellent approximation by the classical
configuration energy\cite{bonsall}.
Because most of the results presented here --- specifically,
low-temperature configurations and depinning fields --- are
essentially static properties of the electrons, they should
be applicable in the presence of strong enough magnetic
fields that exchange effects may be ignored.

Many investigations of systems in the limit $\l_0 \ll a_0$
for the GaAs system showed some signs of behavior associated
with the WC\cite{pudalov}.  In particular, samples with setback ratios
$d/a_0$ of the order 3--4 and larger have been found to
be strongly insulating in the
limit of low temperature\cite{willett1}.  This behavior
is consistent with a crystalline ground state, which is
pinned by an arbitrarily weak impurity potential\cite{fuku},
and hence will not carry current in weak electric fields.
Experiments on samples with still higher setback ratios $d/a_0$,
and remarkably high mobilities, have shown other
signatures of WC behavior\cite{goldman,tsui}.
A depinning\cite{fuku,gruner}
threshold as a function of applied voltage has
been observed\cite{goldman,depin,depin2},
and broadband noise in the current is seen to emerge above
the depinning transition\cite{tsui2}.
(However, narrow-band noise --- often seen in charge
density wave systems\cite{gruner} --- has not been observed\cite{tsui2}.)
Our simulations address this phenomenology, and in particular
attempt to sort out both the qualitative and quantitative changes
in depinning behavior as the strength of the disorder is varied.

As pointed out above, in the low density limit, the
2DEG may be treated as classical.  Using a combination
of molecular dynamics (MD) methods and analysis using
continuum elasticity theory, we have studied the effect that a
disorder potential has on the classical WC.
In this work, we report the results and details of our numerical
simulations.  Our continuum elasticity theory
analysis will presented in detail elsewhere\cite{fertig},
although some of the results of that work will be
explained below.
Some of the results discussed here have
been reported previously\cite{cha}.  The questions we wish
to address are: (1) Is there a qualitative difference
among the ground states of samples subject to different
levels of disorder, or are they the same,
with different correlation lengths?
(2) How does the depinning threshold electric
field vary with the strength of disorder?
(3) How does the current flow when the crystal is depinned,
and in what situations might one expect to observe
narrow band and/or broadband noise?

To be specific, we focus
on modulation-doped semiconductors as our model system,
and control the strength of disorder by varying the ratio
$d/a_0$.  Experimentally, this may be controlled either
by fabricating several samples with different values
of $d$, or by varying the density of electrons in a single
sample (for example, via a gate geometry).
Our study is also directly applicable to studies of electrons
on thin helium films, where disorder may be introduced by using
a glass slide substrate\cite{jiang} as well as to charged polystyrene
sphere systems suspended in water\cite{murray}.
We focus on the extent
of order in typical low temperature states, which are generated
using a simulated annealing method.  We will present
evidence that there is a zero-temperature phase transition
in this system as a function of $d/a_0$, from a state
with power law behavior in the orientational correlation
function to one with an exponential falloff, as $d$ is
decreased.  We call the former state a hexatic glass,
and the latter an isotropic glass.
This transition suggests that there is not just a quantitative,
but rather {\it qualitative} difference among the zero
temperature states of electron layers subject to different levels
of disorder.

We also study how current flows at low
temperature for the different types of ground states,
using our MD simulations, in the presence of an electric
field.  We compute the depinning field $E_{th}$ as a function
of $d/a_0$, and there find direct evidence of the phase transition.
To our knowledge, this is the first suggestion that
depinning properties of a solid
may be used to probe a structural phase transition.
We will demonstrate a striking difference in the
current flow patterns of the two cases.  For the hexatic glass,
we find an essentially elastic flow of the electrons
just above the depinning transition;
i.e., all the electrons participate in the conduction,
and maintain roughly a constant position with respect
to the moving center of mass.  In the isotropic
glass, we find a more plastic flow, in which the crystal
``tears''; in the most
strongly disordered cases, large patches of electrons may be completely
immobilized, while the rest of the electrons carry the current.
Some consequences of the motion for the noise spectrum
will also be discussed.

This article is organized as follows.  For readers interested
just in our results, Section~\ref{sec:model} describes our model, and
then discusses our principle findings.
Readers interested in the details of our calculations
may find our molecular dynamics and simulated annealing
methods described in Section~\ref{sec:method}.
Section~\ref{sec:threshold}
describes the calculations of depinning fields and current patterns,
and we conclude with a summary in Section~\ref{sec:summary}.

\section{Model and Results}
\label{sec:model}

Our model system for the disordered WC is motivated directly
by the modulation doped systems described in the Introduction.
We consider two planes of charges (Fig.~\ref{fig:model}).
One contains $N$ electrons that are free to move within the plane
in response to any forces exerted upon them.
The other contains $N$ positively
charged ions which are quenched; i.e., fixed in position.
These ions are placed in the plane completely randomly,
with no correlation among the ion positions.
The ion layer is setback from the electron layer by a
distance $d$, which controls the strength of the
disorder.  The electrons and ions interact via the
usual Coulomb potential, with a dielectric screening
constant $\kappa=13$, as would be appropriate for
a GaAs system.
We focus on the low temperature states of the electrons
using a simulated annealing method, which will be described
in detail in Section~\ref{sec:method}.
We have studied system sizes up to $N$=3200 electrons
(and 3200 ions), and employ periodic boundary
conditions.  Our electron density in these simulations
is taken to be $5.7 \times 10^{10} cm^{-2}$
We study both the low temperature
configurations of the electrons,
and their response (i.e., motion) when an electric field
is turned on.  The latter is studied using direct molecular
dynamics (MD) simulations of the classical motion of the
electrons; this is described in detail in Section~\ref{sec:threshold}.

The ground states that we find, for {\it arbitrarily large}
values of $d$, are never crystalline: the positional
correlation function falls off {\it exponentially};
i.e., there is short-range positional order.  Such
states are most properly classified as glasses.
It is well-known that any
coupling of randomness to an otherwise perfect crystal
will destroy the long-range positional order\cite{imry}
normally associated with a crystal
in two dimensions.  However, at a minimum
this occurs because of smooth fluctuations that are forced
into the crystal to optimize the total energy (including
coupling to the disorder).  These may be described by a
displacement field $\vec{u}(\vec{R})$, where $\lbrace \vec{R} \rbrace$
are the lattice sites of the perfect crystal, which varies
slowly on the length scale $a_0$.   Such a state
has a {\it power-law} decay in the positional correlation
function\cite{mermin,dzyubenko}, which is sometimes characterized
as quasi-long range order.  Because of this
property, such a state is usually called
a crystal\cite{strandburg}, in spite of the fact that this state is not
crystalline in the sense of classical crystallography.

The states we find, however, are never of this form, regardless
of how weak the disorder (i.e., how large $d$) is.
The reason is that isolated {\it dislocations} become included in the
ground state configuration, which cannot be characterized
by smooth displacements from a perfect crystal.
(A dislocation is a site at which a line of electrons along
a principal axis of the crystal comes to an end.)
This type of defect spoils the quasi-long range order
associated with the ``crystal'' state.

Fig.~\ref{fig:config} illustrates several of the low-temperature states
we have generated using our MD method, for different values of $d/a_0$.
For large values of this parameter [Fig.~\ref{fig:config}(a)],
isolated dislocations
(denoted as bound pairs of $+$ and $\times$ in the figures)
may clearly be seen in the configuration.  As $d$ is further
increased, we have found that the density of dislocations
decreases, until the average spacing between them exceeds
our sample size.  It is interesting to note that grain
boundaries do not appear in our configurations.  Some of the
present literature on the WC has assumed that the effect
of disorder is to introduce well-ordered microdomains, separated by
sharp boundaries.  This would be reflected
in our configuration if the dislocations collected together
to form grain boundaries\cite{nabarro,fisher}.  We have found
that this is not energetically favorable.

As the ratio $d/a_0$ is decreased [Figs.~\ref{fig:config}(b) and (c)],
we have found that the ground state
of the WC undergoes a zero-temperature phase transition, from
a ``hexatic glass'' to an ``isotropic glass''.
Both states must be characterized as glasses, because there
is only short-range order in their positional correlation
functions.
The difference between these states may be understood in
terms of defects called {\it disclinations}\cite{nabarro,nelson1}.
These are
lattice points which have the incorrect number of nearest
neighbors.  In a perfect WC, which has
a triangular lattice as its ground state\cite{bonsall},
disclination points may have (for example) five or seven
nearest neighbors, whereas the perfect lattice points
have six.
A charge $s$ may be assigned to the disclinations, so that
$s+6$ is the number of nearest neighbors surrounding
the disclination.  In the hexatic glass state, disclinations
are present in tightly bound pairs (``neutral'', in terms
of disclination charge,) separated by a distance of order of a lattice
constant.  Such a bound pair is equivalent to a dislocation.
One never finds isolated disclinations in this state.
As the ratio $d/a_0$ is decreased, one may see disclination
pairs which are separated by more than a single lattice
constant, but are clearly still bound together.  Above
the critical setback distance, one can find isolated disclinations
in the system.

As in the case of isolated dislocations, isolated disclinations
tend to spoil correlations.  The relevant correlation function
for this disclination-unbinding transition measures
the {\it orientational} order in the system.
In the absence of isolated disclinations, the orientational
correlation function falls off only as a power law
(quasi-long range order); when
they are present, it falls off exponentially (short range order).
Fig.~\ref{fig:correlation} illustrates the orientational correlation functions
for several values of $d/a_0$.
(Precise definitions of the positional and orientational
correlation functions will be given in Section~\ref{sec:method} below.)
We estimate from this
that the transition between the states occurs
near $d_c/a_0 = 1.15$. It must be noted that our simulations
include neither the finite thickness of the layer, nor the
finite value of the magnetic length $l_0$ when a magnetic
field is present.  Both these effects tend to soften the
electron-electron interaction relative to the electron-ion
interaction, so we expect the actual value of $d_c$ to
be somewhat higher than this value.  Nevertheless, it would
be useful and interesting to look for this transition in
real samples --- either in heterostructures or on He film
systems --- as a function of $d/a_0$.  A novel way of
detecting the transition by measuring the depinning field
is described below.

Readers familiar with the Kosterlitz-Thouless-Nelson-Halperin-Young
(KTHNY)
theory of two-dimensional melting\cite{nelson1,strandburg,kost}
will recognize the phenomenology of this transition.
In the KTHNY theory, a two-dimensional crystal in the absence
of quenched disorder but at finite temperatures, undergoes
two phase transitions as temperature is raised.  At the lowest
temperatures, thermally activated
dislocations are bound together into pairs
with equal and opposite Burger's vectors\cite{c1}.
This state is a crystal in the sense described above;
there is a power law decay in the positional correlation
function.  Furthermore there is long-range order
in the orientational correlation function.
Above the
Kosterlitz-Thouless melting temperature\cite{kost}, these pairs
become unbound, and the crystal melts into a ``hexatic'' phase,
characterized by short range positional order and quasi-long
range orientational order.  There are no isolated disclinations
in this state.
Finally, at a second higher temperature,
the disclinations that make up the dislocations themselves
unbind, and one finds both isolated dislocations and
disclinations in typical configurations.  Both the positional
and the orientational correlation functions are short ranged,
so that this state may be appropriately characterized as
a liquid.

Clearly, the zero-temperature behavior
of the electron solid in the presence of quenched disorder (i.e.,
the inherent disorder due to the random locations of the
dopant ions) is partially analogous to this.  In particular,
we see an analog of the disclination unbinding transition;
however, the
dislocations never pair together, even in the very weakly
disordered case.
It has been argued previously, based on a
renormalization group analysis of a model
closely related to ours, that a crystal is
unstable with respect to the formation of free
dislocations in the presence of arbitrarily
weak quenched disorder\cite{nelson2}.
Both these results ---
the absence of the crystal state, and the
disclination unbinding transition ---
may be understood
from a continuum elasticity theory
model of the electron crystal.
The details of this analysis will be given elsewhere\cite{fertig}.

It is interesting and instructive to see how the impurity driven
disclination unbinding transition is analogous
to the temperature driven Kosterlitz-Thouless
transition\cite{kost}.
Here we
present only the result; details will be presented
elsewhere\cite{fertig}. When one
takes into account screening
by dislocations\cite{c2}, it may be shown that
the energy to create an isolated disclination (along with
its screening cloud) has the form $E_1 = E_c \log (A/a_0^2)$,
where $A$ is the system area and $E_c$ is the core
energy of a dislocation.
However, there is also an energy of interaction $E_2$
between the disclination and the free dislocations
created by the impurities.
If we consider an ensemble of disorder realizations,
there will be a
probability distribution $P(E_2)$ for $E_2$ to take on
a particular value.
The distribution
of these energies we find\cite{fertig,c3} to take
the form $\exp\Bigl[{-E_2^2/4\pi\rho_b a_0^2 E_c^2 \log (A/a_0^2)}
\Bigr]$,
where $\rho_b$ is the dislocation density (in the
absence of the disclination).
The probability that it will
be energetically favorable to create a disclination
at a given site ($E_1+E_2 < 0$) thus scales as
$A^{-1/4 \pi \rho_b a_0^2}$.  Noting that the number
of available sites to create a disclination scales as $A$,
the total number of sites in a sample for which it is
favorable to create a disclination
scales as $A^{1-1/4 \pi \rho_b a_0^2}$, and
is non-vanishing
in the infinite size limit if
$$\rho_b a_0^2 > {1 \over {4\pi}} \approx 1/14.$$

This is the analog of the original Kosterlitz-Thouless result\cite{kost}.
We have found in our simulations that the transition appears
to occur at slightly lower dislocation densities,
around $\rho_b a_0^2 \approx 1/(20 \pm 2)$.

Beyond the ground state structure of the WC in the presence of
quenched disorder, we have also investigated the depinning
properties of this system.  Fig.~\ref{fig:trajectory} illustrates the
trajectories of the electrons over a finite time interval
for different values of $d/a_0$,
for electric fields just above the depinning threshold.
Fig.~\ref{fig:trajectory}(a) shows
these trajectories for $d/a_0 = 1.5$.  The motion of the
electrons is quite uniform, and may be described as an
elastic flow.  It is important to note, however, that in
the {\it time} domain, this flow is not as uniform as it
appears in this figure:
different patches of the WC actually move at different moments,
so the motion is more of a ``creep'' than a flow.
However, when averaged over a long enough time, the
flow is uniform.

As the disorder increases [Fig.~\ref{fig:trajectory}(b)],
this creeping motion becomes more pronounced,
with the waiting time for motion of certain patches
sometimes becoming quite long.  There is also an interesting
and somewhat surprising phenomenon
that becomes important as we pass from the
hexatic to the isotropic glass:
the direction of motion is correlated with the local
crystal axis directions, so that the the local velocity is not exactly
along the direction of the applied electric field.  Because
the system does not possess long-range orientational order,
this means that different regions will slide in different
directions, and in particular will at some locations
``crash'' into one another.  For such grains, there
may be long waiting times before the electrons at the
regions where these merge can rearrange themselves and
continue on.  The rearrangement is often accompanied by
the temporary formation of defects (dislocations or disclinations)
which appear in the process of rearrangement.  These
regions clearly represent bottlenecks in the flow of
the electrons, since crystal must become highly strained
in order for the different flow directions to resolve
themselves.  Because of the rearrangement of the lattice
in the process of sliding, this flow must be described as
plastic.  The importance of plastic flow in determining depinning
properties has been emphasized recently for CDW systems\cite{copp}.

At still stronger disorder strengths [Fig.~\ref{fig:trajectory}(c)],
the creep behavior completely dominates the motion of the electrons.
In this limit, the wait time for some patches to move is
longer than our simulation times; i.e., certain regions of
the crystal never move at all.  This means the solid ``tears''
when it depins.  The flow in this limit appears channel-like,
with only certain regions participating in carrying the
current.  Such channel-like motion has been observed
previously in studies of the depinning properties of (and
diffusive motion in)
flux lattices in thin superconducting films\cite{berlinsky}.
Another important aspect
of the motion that is not obvious from the trajectory plots
is that the channels do not remain constant as a function
of time: different channels appear to open and close as the
simulations progress.  Thus, most of the particles do participate
in the current, although at any given moment only a small
percentage of them have a significant velocity.

Another interesting result of our work is the behavior of the
threshold electric field $E_{th}$ for depinning as a function of
$d/a_0$.  We find that the depinning field scales exponentially
with the setback distance, as may be seen in Fig.~\ref{fig:efield}.
This effect may be understood if one assumes that the pinning
is roughly determined by the strength of the electron-ion potential
interacting with a perfect WC.  This interaction scales
approximately as\cite{cha,fertig2} $e^{-Gd}$, where $G$
is a typical primitive reciprocal lattice vector magnitude,
of order $2\pi / a_0$.  It should be noted, however, that
that the defect structures induced
by the disorder are also important in determining the strength
of the depinning field.  Configurations generated by relaxing
an initially perfect crystal in the presence of the disorder
potential, which does not allow isolated dislocations
or disclinations to form in the resulting state, tend to have significantly
smaller depinning fields than those containing these defects.
The great sensitivity of $E_{th}$ to $d/a_0$ may potentially
explain the rather large disparity found in measurements
of the depinning field in different samples\cite{depin2}.

A very unusual property of $E_{th}$ is its behavior in
the vicinity of the transition from the hexatic to the
isotropic glass.  One can see a break in the curve in the
vicinity of $d=1.15a_0$, which we associate with the
transition between states.
Physically, we associate
this break with the change in the orientational correlation
functions in the vicinity of the transition.  Since the
flow patterns of the electrons as they depin follow the
local orientational axes of the crystal, the bottlenecks
in the flow occur where the electron motion must change directions,
as described above.  The number of such bottlenecks clearly
will proliferate rapidly as the orientational order
changes from exponential behavior to power law behavior.
This explains the increase in slope in Fig.~\ref{fig:efield}
as $d/a_0$ drops below about 1.15.
We note that the break is somewhat
rounded, which we associate with the finite number
of particles in our simulations.
We have found that the break becomes slightly more
rounded for smaller systems (although our error bars
also increase for these simulations.)
We believe it is likely
that this break will become sharper in the limit of
infinite system sizes.  An observation of such behavior
in the depinning field would give direct evidence of
a transition between the hexatic and isotropic glass
ground states.

Finally, it is interesting to speculate what effect the creeping motion
has on the noise spectrum in the current.
In the simplest case of a sliding CDW, one expects
the noise spectrum to have a narrow band component,
corresponding to the sliding of the (periodically arranged)
electrons over
the impurities\cite{gruner}.  The frequency of this noise
component is proportional to the average velocity $v$ of
the electrons.

Certainly for the case of our more disordered samples, this
picture breaks down.  The flow of electrons when they are depinned
is plastic rather than elastic, so that the electrons do
not preserve a local crystal symmetry particularly well
as they are sliding.  Furthermore, the opening and closing
of various channels through the system introduces many time
scales, contributing to a severe broadening of the noise spectrum.

Even at larger values of $d/a_0$, we do not observe
a perfectly uniform motion of the electrons; we still
observe a kind of creeping, jerky motion.
Near the threshold voltage, the
average time for the crystal to move one lattice constant,
$a_0/v$, will become longer than the average time an individual electron
remains stationary.  In this circumstance
one does not expect to observe narrow band noise.  Thus,
narrow band noise should only be expected at voltages
well above the threshold voltage for sliding.  We have
observed narrow band noise in this circumstance in
our simulations.
Experimentally, however, it must be noted that
finite currents can ``heat'' the electrons,
which would lead to a melting of the crystal\cite{depin2,c4}.
At present it is unclear whether the currents required for
the narrow band noise to become visible over the noise
introduced by the creeping motion are low enough to
avoid such effects in realistic samples.
More detailed investigations of
this issue are currently underway.

\section{Molecular Dynamics and Simulated Annealing Method}
\label{sec:method}

Molecular dynamics (MD) is a powerful and well-known method for
studying the motion of classical particles\cite{hockney}.  The central
idea is the discretization of time, so that Newton's law
$\vec{F}_i(t) = m{{d\vec{v}_i(t)} \over {dt}}$ and the kinematic relation
${{d\vec{r}_i(t)} \over {dt}}=\vec{v}_i$ may be rewritten as difference
equations.  Here, $\vec{F}_i(t), ~\vec{v}_i(t)$, and
$\vec{r}_i(t)$ are respectively the total force acting on
particle $i$, the velocity of particle, and its position,
at time $t$.  The forces on a given particle in
our simulations come from one
of three sources: (1) Interactions with other electrons,
$$
\vec{F}_i^{ee} = \mathop{{\sum}'}_{\vec{r}_j}
{{e^2} \over {\kappa |\vec{r}_i-\vec{r}_j|^3}}
(\vec{r}_i-\vec{r}_j),
$$
where the prime in the sum indicates that $\vec{r}_i$
should not be included in the sum over $\vec{r}_j$, $e$
is the electronic charge, and $\kappa$ is the dielectric
constant of the host medium.
(2) Interactions with the ions,
$$
\vec{F}_i^{ei} = -\sum_{\vec{R}_j}
{{e^2} \over {\kappa (|\vec{r}_i-\vec{R}_j|^2 + d^2)^{3/2}}}
(\vec{r}_i-\vec{R}_j),
$$
where the $\vec{R}_j$'s are the ion positions in their setback
plane.  (3) Interactions with external forces (e.g., an external
electric field.)  We assume periodic boundary conditions,
so that our system size is formally infinite, although it
is periodic with a large but finite number of particles
in each unit cell.  The number $N$ of electrons (and ions)
is taken to be 3200 in all the simulations reported
here, except where specifically stated otherwise.
Because of the long range nature of the
Coulomb interaction, the forces in (1) and (2) have to be
computed using the Ewald sum technique\cite{bonsall},
which constitutes the bottleneck in our computations.

Part of our simulations require us to work at finite temperatures,
which can be accomplished in one of two ways.  In the
microcanonical ensemble, we fix the total kinetic energy
of the system to $Nk_BT$, where $k_B$ is Boltzmann's
constant, and $T$ is the temperature.  This is accomplished
by rescaling the velocities by the rule
$\vec{v}_i(t) \rightarrow \alpha \vec{v}_i(t)$,
such that $\sum_i {1 \over 2} m v_i^2 \equiv Nk_BT$
at every time step.  The temperature of the system
can be checked by examining the changes in the internal
potential energy per particle of the system
$U(t)/N$, where
$$
U={1 \over 2} \sum_{i\ne j} {{e^2} \over
{\kappa |\vec{r}_i-\vec{r}_j|} }
-
\sum_{i,j}{{e^2} \over
{\kappa (|\vec{r}_i-\vec{R}_j|^2+d^2)^{1/2}} }
+
{1 \over 2} \sum_{i\ne j} {{e^2} \over
{\kappa |\vec{R}_i-\vec{R}_j|} }.
$$
Note that the sums are over all the particles and ions in all
the unit cells, and hence is a sum over an infinite number of
particles.  The last term in $U$ is independent of the
electron positions and so does not contribute to its changes
(the ions positions are quenched, and do not move in our
simulations), but it is necessary to keep this term in order to
get a finite result.  When the system is in equilibrium, we
find that at low temperatures
$\overline{U_{T_1}(t)}-\overline{U_{T_2}(t)}=N k_B (T_1-T_2)$,
where $U_T(t)$ is the configurational energy at temperature $T$
at time $t$ and the overbars represent time averages.
Thus, classical equipartition of the energy satisfied when our
simulations are in thermal equilibrium.

A second method for simulating the system at
finite temperatures is to use the Langevin equation\cite{hockney,balescu}.
In this method, each electron $i$ is subjected to a white noise
random force $\vec{\xi}_i(t)$ and a viscous force
$F^{\eta}=-\eta \vec{v}_i$.  The random force
is connected to the temperature via the correlation function
$\overline{\xi_i^{\alpha}(t) \xi_j^{\beta}(t^{\prime})} =
-2\eta k_B T \delta(t-t^{\prime})\delta_{ij}\delta_{\alpha \beta}$,
where $\alpha,\beta~=~x,y$.  We have found that this method
gives results similar to those obtained by the microcanonical
ensemble.  We will report in detail below our results
as obtained in the latter method.

To obtain typical low energy configurations, we employ
a simulated annealing method.  Our procedure begins by
assigning to the electrons random positions in the plane,
and velocities according to a Gaussian distribution centered
around our chosen temperature.  The initial temperature is
taken to be large enough that the electronic state is a liquid.
The positions and velocities of the electrons are updated
for several thousand timesteps, in order to assure that
the system is in thermal equilibrium.  Once thermal equilibrium
appears to have set in, the temperature is lowered by a small
amount $\Delta T$, by rescaling all the velocities.  This
temporarily puts the system out of equilibrium, and it is
necessary to update the system for several thousand
timesteps to equilibrate it.  This process is repeated until
the system approaches the melting temperature.

The melting transition in the absence of quenched disorder
for two-dimensional classical electron systems
has been studied via numerical methods by many
authors\cite{strandburg}. The temperature of this
system may be expressed in terms of the single unitless
parameter $\Gamma = \sqrt{\pi \rho} e^2 /\kappa k_BT$, which is
the ratio of the average potential energy to the average
kinetic energy of the system.  In this expression, $\rho$
is the two-dimensional electron density.  It is generally
believed, based on Monte Carlo simulations\cite{gann,strandburg}
that the electron crystal melts in the vicinity
of $\Gamma=130$.  Thus, the freezing temperature for our
system should occur in the vicinity of 418mK.

It is extremely
important that in the annealing process the system spends
a large number of time steps near and below this temperature in
order to have a well-equilibrated system.  The reason is that,
once the system begins to freeze, defects will migrate very slowly.
The energy barriers required to create or eliminate defects
become quite high well below the freezing temperature,
so that long annealing times are necessary to eliminate defects that
should only be present at higher temperature.
Thus, it is necessary to equilibrate
the system with a sufficient number of steps when temperature is
below the melting point.
We have checked this
by slowing our annealing process (i.e., decreasing $\Delta T$ and
increasing the number of time steps at each temperature)
until the number
of defects in the frozen state for a given impurity configuration
is essentially unchanged.  Another non-trivial check that our
annealing process equilibrates the system is to run it in the
absence of any impurities (i.e., with a uniform neutralizing
background), in which case the ground state should be defect-free.
We have found that this is the case in most situations\cite{c5}.

The type of defects that are tracked in our simulations are
the number of disclinations --- i.e., the number of electrons
with the incorrect number of nearest neighbors.  As discussed
in Section~\ref{sec:model}, a dislocation is equivalent to a bound pair
of disclinations, so implicitly this keeps track of the
dislocations as well.  The defects are located using
the Voronoi polygon method\cite{polygon}, which is a numerical
method by which the Wigner-Seitz cell around each electron
may be constructed; the number of nearest neighbors is then
equal to the number of sides of the cell.

The results presented here are for very low temperatures
compared to the melting temperature, typically of the order
20mK.  Finally, it should be noted that because of the finite
annealing time, one cannot expect,
if the temperature were lowered precisely to zero, that the
resulting state
found in our simulation would be the global ground state of the
system.  It clearly will be a low energy metastable state, and
there is no reason to believe that it will be qualitatively
different than the ground state.  Thus, for the properties
we are interested in
--- correlation functions,
number of defects, depinning fields ---
the annealed state should give qualitatively the same results as
the true ground state of the system.

Some typical configurations are illustrated in Fig.~\ref{fig:config},
for various setback distances.
As was mentioned in Section~\ref{sec:model}, we have found
that the system is always unstable against the
formation of dislocations for any level of disorder (i.e., any
value of $d/a_0$), and that isolated disclination starts
to appear in the vicinity of $d/a_0 \le 1.15.$  The
dislocations may be seen to appear in regions where there
is a gradient in the electron density --- i.e., there is a local
change in the lattice constant of the WC.  Physically, this
may be understood in terms of the long-wavelength fluctuations
in the positively charge ionic background.
It is convenient to reexpress the ionic charge as a positively
charged, nonuniform neutralizing background that is
in the same plane as the electrons\cite{fertig2}.  The long wavelength
behavior of WC will be essentially determined by charge
neutrality: since maintaining charge imbalances over long
distances is prohibitively expensive energetically, the
WC will arrange itself in such a way as to neutralize
this effective, spatially fluctuating positive charge.

This is essentially the driving force for the creation
of dislocations.  To see this, consider a region $\Omega$ of
size scale $\xi$ with a slightly larger neutralizing
background density than the average, $\rho_0+\delta\rho.$
The WC must raise its local density to neutralize
this fluctuation.  If this is done by smooth displacements
from a perfect crystal, the resulting configuration will
be strained in the vicinity of the region $\Omega$,
with an energy cost scaling as $\xi^2$.  The WC can lower its
energy by introducing
dislocations near the boundary of $\Omega$.  These will
allow lines of charge to be added or removed from $\Omega$,
so that the crystal is essentially unstrained both inside
and outside $\Omega$.  The energy cost to introduce these
dislocations\cite{cha,fertig} scales as $\xi \log{\xi}$;
thus, for large enough $\xi$ --- i.e., for long wavelength
fluctuations --- it is energetically favorable to create
highly separated dislocations.
This is illustrated explicitly in Fig.~6.
We have taken a small-sized system --- 480 particles ---
and introduced a circular region $\Omega$ at the center of our
unit cell with density $\rho_0 + \delta \rho, \delta \rho= {6
\over 25} \rho_0$.
In Fig.~6(a), we obtain a low-temperature configuration by a
direct relaxation method\cite{fisher},
which does not allow defects to be introduced in the final
configuration.  The strain in the lattice outside $\Omega$
is quite apparent.  In Fig.~6(b), we have used our simulated
annealing method to find a low energy configuration.
As may be seen, the strain has been relieved, at the expense
of introducing several dislocations near the boundary.
The configuarion in Fig.~6(b) is lower in energy than that of
6(a), by approximately 90 mK.

In our model, the correlation function for such
fluctuations in the effective in-plane neutralizing background
may be shown to
be $\langle \rho_n(\vec{q}) \rho_n(-\vec{q}) \rangle= A \rho_0
e^{-2qd}$, where $\langle\cdot\cdot\cdot\rangle$ represents
a disorder average,
$\rho_n(\vec{q})$ is the Fourier transform
of the effective in-plane neutralizing background charge
density, $\rho_0 = N/A$ is the average density of ions,
and $A$ is the system area.  Note that as $d$ is increased,
it is the {\it short} wavelength fluctuations that are
supressed.  However, since dislocations arise due to
long wavelength fluctuations in the disorder, we do not
expect that large values of $d$ will suppress them.
We have observed this behavior in our simulations.

As discussed in Section~\ref{sec:model}, we have calculated both
positional and orientational correlation functions
for our low-temperature states.
We conclude this section by giving precise definitions
of these correlation functions\cite{Nelson1982}.
The positional correlation function is given by
$$
g(r) =
{\sum_{i,j}
\delta(r-|\vec r_i-\vec r_j|)
{1 \over 6} \sum_{\vec G} e^{i (\vec r_i - \vec r_j)\cdot \vec G}
\over
\sum_{i,j}
\delta(r-|\vec r_i-\vec r_j|)},
$$
where $\vec G$ is a reciprocal lattice vector with the orientation
which gives a peak of the structure factor.
In practice, the $\delta$-function must be broadened
so that it may be handled numerically.  An example of the
positional correlation function is illustrated
in Fig.~\ref{fig:corrt}.  The long distance tail of $g(r)$ falls off
exponentially, indicating that there is only short
range positional order in the system.  As stated in Section~\ref{sec:model},
$g(r)$ has this behavior independent of whether it is in
the hexatic or isotropic phase, i.e., for all values of $d$.

More interesting behavior is apparent in the orientational
correlation function.  In analogy with Ref.~[25],
this is defined as
$$
g_6(r) =
{\sum_{i,j}
\delta(r-|\vec r_i-\vec r_j|)
\psi(\vec r_i)^* \psi(\vec r_j)
\over
\sum_{i,j}
\delta(r-|\vec r_i-\vec r_j|)},
$$
where
$
\psi(\vec{r}) = (1/n_c)\sum_{\alpha \in \{n.n.\}}
e^{6i\theta_\alpha(\vec r)}
$
for a electron located at $\vec r$ with the bond angle
$\theta_\alpha(\vec r)$ with respect to the $x$ axis
to the $\alpha$th nearest neighbor,
summed over $n_c$ nearest neighbors determined by
the Voronoi polygon method\cite{polygon}.

Examples of $g_6$ are presented in Fig.~\ref{fig:correlation},
where the change in behavior
in going from the isotropic to the hexatic phase is apparent.
In particular, in the former case $g_6$ is short-ranged, while in
the latter we expect a power-law behavior at long distances\cite{nelson1}.
Our simulations are consistent with this, although because of
the finite size of our simulations it is difficult to distinguish
between short-range order with a very long correlation length,
and power law behavior.
However, direct estimates of the correlation length
from $g_6$ in the interval
$1.1 < d/a_0 < 1.2$ show that it increases very rapidly
in the range, as would be expected for a critical phenomenon.
We note that statistics for both
$g$ and $g_6$ can in principle be improved by averaging
over impurity configurations.  Unfortunately,
this is not practical, as the large size systems required ($\sim 3200$
particles) to reliably measure the correlation functions in
a single sample prohibit a large number of repititions.

Another way of displaying the onset of hexatic order is to
measure the structure factor.  This is defined in the usual
way as $|S(\vec{q})| =
{|\sum_i e^{i\vec{q} \cdot \vec{r}_i}|}$, and is
illustrated in Fig.~\ref{fig:sfactor} for several setback distances.
For small setback distances $|S(\vec{q})|$ is essentially circularly
symmetric; but as $d$ is increased, one can see six well
defined peaks developing for $d/a_0 \ge 1.2$.  It should be noted
that these are not Bragg peaks, in the sense that they do
not become sharp in the large size limit; they instead
remain as a modulation in an increasingly strong
background\cite{nelson1,Nelson1982}.
Nevertheless, the presence of this modulation indicates the
existence of some orientational order in the hexatic glass
phase.

\section{Depinning Thresholds and Currents}
\label{sec:threshold}

In this section, we discuss the numerical method used to compute
the threshold electric field above which the WC depins
and the path the current follows above this threshold.
Conceptually, the simplest way to do this is to
begin with an annealed system, prepared as described
in Section~\ref{sec:method}, and then simulate
a uniform electric field by subjecting all the particles to
a constant force.  In practice, however, this turns out
to be computationally quite inefficient.  The reason is that
when the electric field is  stepped up by
an amount $\Delta E$, this represents a large, non-random ``kick'',
after which the system must re-equilibrate.  These equilibration
times turn out to be quite long, typically on the order of
5000 time steps for 3200 particles.  If the electric field
step $\Delta E$ is taken to be too large, or if the equilibration
time is too short, one can reach a metastable state in
which current flows for several thousand time steps, but
then may abruptly stop when the random motion of
the electrons around the moving center of mass equilibrates.

Fig.~\ref{timecurrent} illustrates the current for such a sequence of
electric field increments, in which the net electric field
remains below the threshold.  Occasionally, after a
particular increment, one can see very large amplitude
oscillations, far above that expected from simple random
motion (e.g., first 2000 timesteps in Fig.~\ref{timecurrent} )  Such
large oscillations occur if the fluctuating center
of mass happens to be moving in the same direction as the
uniform force applied to the electrons at the moment
it is incremented.
Apparently this can set into motion very long-lived
phonon modes, and one must wait long times for
the energy in this mode to redistribute itself
among the other phonon modes -- i.e., for the
phonon to decay.
Such a process occured at the moment the
electric field was turned on in Fig.~\ref{timecurrent}
(27000{\it th} timestep),
and one can see that the resulting non-equlibrium
oscillations did not settle down even after 4000
timesteps.

A much more numerically efficient method for determining
the depinning threshold using MD was developed by
Brass and coworkers\cite{brass}.  In this technique, one
begins with an annealed low temperature state, and then
shifts all the particles by a small amount ($0.01 a_0$ in
our simulations).  The MD method is then used to
reequilibrate the system, {\it keeping the center
of mass of the electrons fixed in position}.  This
last requirement is enforced by adding a small correction
to the velocities of all the electrons at each update,
$\Delta v(t)$, so that the center of mass velocity remains
precisely zero at every timestep.  The pinning force,
defined as the average force acting on an electron
due only to the impurities, is then measured.  The sequence
is then repeated, until the total shift in position is
of the order $1-3a_0$.

The great numerical savings in this scheme is that the number
of timesteps required to equilibrate after a center of
mass shift is quite small, only around 200 for 3200 particles.
Thus, the time required to find the threshold field in this
manner is approximately an order of magnitude smaller than
finding it by direct simulation of the electric field.
Fig.~\ref{pinf} illustrates the pinning force (represented as an
electric field, by dividing out the electric charge)
in a sequence of
center of mass shifts.  As can be seen, the pinning force
rises monotonically (except for small fluctuations) to
some maximum, and then abruptly drops.  We associate this
maximum with the depinning electric field.  The trajectories
of the particles for more disordered samples
support this interpretation: one can see large rearrangements
of the electrons (``channel-like flow'') as the pinning
force is decreasing.  The sequence of steps gives several
peaks of roughly the same order of magnitude, and the largest
of these is chosen as the threshold electric field, $E_{th}$.
The variance
among the peak heights gives a measure of the uncertainty
in this quantity.

We note that we have compared $E_{th}$ for
several samples, using both direct simulation of an electric
field and the shift method.  The agreement between the
thresholds found by these two methods is quite good.
The qualitative behavior of the electronic
motion for various levels of disorder found via the two methods
-- i.e., elastic flow vs. plastic flow or channel-like
flow -- is also very similar.

Another detail that requires attention for the highly
disordered samples is the fact that
the disorder potential in general will favor certain
regions for the channels through which the current
may flow.  Periodic boundary conditions can actually
mask this effect, particularly if the channels are narrow,
because in general a channel
enters and exits at different points along the boundaries.
This can lead to an overestimate of the pinning potential,
since the final current channel may have to run through the
sample many times before closing in on itself.
Such narrow channel flow (``string-like motion'') has
been observed in MD simulations of vortices in thin
superconducting films, subject to a high degree
of disorder\cite{berlinsky,brass}.
To overcome this potential difficulty, we have used samples
which have ion distributions that are mirror-imaged
across the $y$-axis through the center of the sample.
Thus, if the disorder potential favors relatively narrow
channels, the crystal
will not be pinned simply because the current has nowhere to go.

In practice, we have found that this consideration only affects
our results for the most extremely disordered samples.
Most of the electrons do eventually participate
in the current, although they do not in general move at the
same time.  Even in the more disordered samples, where the flow
is channel-like, the current channels are actually quite
broad compared to what has been found in vortex
systems\cite{berlinsky,brass}.  Furthermore, different
channels open and close with time, so that
over the simulation time, current flows through most
parts of the sample (although some patches do remain pinned
thoughout the simulation.)
Thus, the disorder does
not pick out a small region over which most of the current
flows for the WC, except possible at exceptionally small
values of $d/a_0$.

Our results for the depinning
field at various setback distances for a system of 3200
particles are shown in Fig.~\ref{fig:efield}, and were discussed in
Section~\ref{sec:model}.  As was noted, there is a noticeable break
in $E_{th}$ vs. $d$ in the range $d/a_0=1.1-1.2$,
precisely where we see the transition between the
hexatic and the isotropic glass phases.  This
may be understood in terms of the qualitative behavior
of the electron motion as the disorder is turned up ($d$ decreased),
changing the motion from elastic to plastic, channel-like flow.
In the latter regime, especially near the hexatic-isotropic
transition, there is a tendency for the flow to follow the
local orientational axes of the crystal.  Since the orientational
correlation function changes from quasi-long range to short range
as $d$ is tuned through the transition, the number of changes
in direction that a group of electrons will have to undergo
in order to flow through the crystal increases rapidly once
we enter the isotropic phase.  The changes in direction
represent bottlenecks in the motion, and cause the depinning
field to increase rapidly as $d$ decreases.  An experimental
observation of this change of behavior in $E_{th}$ vs. $d$
would represent a direct and unique way of probing the
transition from the hexatic to the isotropic glasses.
One might also note that the uncertaintly in our results
increases significantly with increasing $d$.  This is because
the number of defects in the system drops quickly in this
limit, so that our results become more sensitive to
finite size effects and fluctuations from one impurity
realization to another.  These uncertainties would certainly
be reduced with studies of significantly larger systems,
although this seems impractical at present from a numerical
point of view.

Finally, we compare our results with recent calculations
by Ruzin et al.\cite{ruzin} on pinning due to stray
charged impurity ions that may lie relatively close to
(i.e., a distance on the order of $a_0$ away from)
the electron plane.  This is relevant to modulation
doped semiconductor systems, in which a perfectly pure
setback layer is in practice not possible to fabricate;
small densities of donors and acceptors inevitably become
incorporated in this layer.  While detailed comparisons
are difficult because little is known about the
density and distribution of these unintentional dopants,
we can speculate as to how they affect our results.

We note first that the state (i.e., hexatic or
isotropic) of the system is unlikely to
be affected by these impurities, since at such
sparse densities they induce point defects (i.e., interstitials
and vacancies)
rather than extended defects, such as dislocations and
disclinations.  However,
it is clear that for large enough setback distances
$d$, the impurities in the volume of the setback
layer will dominate the pinning properties.  Using
a density of $10^{13}$ cm$^{-3}$ positively
charged donors\cite{c6}, and assuming that
all these impurities lie at a distance from the
electron plane that maximizes the pinning, we obtain
an upper bound of $10^{-4}e/\kappa a_0^2$ for the
pinning threshold due to these stray impurities.
This is a very conservative estimate, and it should
be noted that the true pinning threshold due to such
impurities will likely be significantly lower than this.
(Indeed,  in experimental systems with the largest
setback ratios reported thus far\cite{goldman,tsui} ($d/a_0 \approx 6$),
the pinning threshold electric field is significantly
smaller than this.)  This estimate
is near the bottom of our Fig.~\ref{fig:efield},
so we do not expect it to affect our results for the
hexatic-isotropic transition.  Furthermore, an extrapolation
of our results (Fig.~\ref{fig:efield}) to large $d$ can give rough
agreement with experimental samples\cite{c7}.
Future
samples with very large values of $d$ may well be
dominated by stray impurities in their pinning properties.
It is difficult to ascertain at this point -- both because
of the uncertainty in the density of these impurities, and
the error bars in our calculations for large values of $d$ due
to the finite size of our system -- their relative importance
in currently available semiconductor systems.
Clearly, more work (both experimental and theoretical)
is required to sort out this issue.

\section{Summary}
\label{sec:summary}

In this article, we have studied using molecular dynamics techniques
the low-temperature configurations and the depinning properties of
the classical electron solid in the presence of disorder.
Our calculations are applicable to both electrons on the
surface of helium, and to the two-dimensional electron gas
in very strong magnetic fields.
We have found evidence for a transition from a hexatic glass
state to an isotropic glass state as a function of the
disorder strength, which is characterized by the appearance
of isolated disclinations, and the simultaneous loss
of quasi-long range orientational order as one passes
from the former to the latter.
This result indicates that the ground states of
2DEG's in modulation doped
semiconductors in the WC regime are {\it qualitatively}
different in the weak and strong disorder regimes.
We have found that
the depinning threshold electric for the system increases
rapidly (exponentially) with decreasing $d$, and that
there is a break in $E_{th}$ vs. $d$ in the vicinity of
the transition between the states.  An observation of
the latter behavior in experiment would represent a
direct probe of the transition.  Finally, we observed
a qualitative change in the motion of the electrons
when they were depinned, from elastic flow in the
very weakly pinned regime, to plastic, channel-like flow
in the strongly pinned regime.

\acknowledgements
The authors thank Steve Girvin and Clayton
Heller for helpful discussions.  This work was supported
by NSF Grant No. DMR92-02255.  HAF also acknowledges an A.P. Sloan
Foundation Fellowship and a Cottrell Scholar Award.

%%%Figure Captions
%%%FIG.1
\begin{figure}
\caption{Model system of $N$ electrons and $N$ quenched positively
charged ions, in planes separated by a distance d.}
\label{fig:model}
\end{figure}

%%%FIG.2
\begin{figure}
\caption{Sample ground state configurations,
for different levels of disorder.
Locations of disclinations are marked by $\times$ for a seven-fold
site, and $+$ for a five-fold site.
Bound pairs of these defects are equivalent to dislocations.
Pictures contain $\approx 1600$
particles;  actual simulations samples contained 3200 particles.
(a) $d/a_0$ = 1.5, (b) $d/a_0$ = 1.2, and (c)$d/a_0$ = 1.0.}
\label{fig:config}
\end{figure}

%%%FIG.3
\begin{figure}
\caption{Orientational correlation functions for various setback
distances.  Different symbols represent data for samples with
different setback distances.  From top to bottom,
$d/a_0 = 2.0.~1.7,~1.5~,1.3~,1.2~,1.1,~1.0,~0.9,$ and 0.8.}
\label{fig:correlation}
\end{figure}

%%%FIG.4
\begin{figure}
\caption{Trajectory plots for 3200 particles
for depinned WC with various levels of disorder.
(a) $d/a_0=1.5$, (b) $d/a_0=1.2$, and (c) $d/a_0=0.9$,
taken over a center of mass shift of
$\Delta X/a_0=0.56, 0.44$, and 0.36, respectively.}
\label{fig:trajectory}
\end{figure}

%%%FIG.5
\begin{figure}
\caption{Depinning threshold electric field, in units of
$E_0=e/\kappa a_0^2$.  Dotted lines are guides to the eye.}
\label{fig:efield}
\end{figure}

%%%FIG.6
\begin{figure}
\caption{At long wavelengths, the electron crystal density must
track the disordered neutralizing background charge density.
(a) Changing density with smooth displacements of an undefective
WC produces a strained crystal in the region $\Omega$ circled by
dotted lines.
(b) Introducing dislocations in the region where the background
density changes by $\delta \rho$ allows the electron density
to match the background density without large regions of
strain.}
\end{figure}

%%%FIG.7
\begin{figure}
\caption{Positional correlation functions for thress different
setbacks.~~~~~~~~~~~~~~~~~~~~~~~~~~~~~~~}
\label{fig:corrt}
\end{figure}

%%%FIG.8
\begin{figure}
\caption{Structure factor $S(\vec{q})$ in the reciprocal
lattice vector space for samples with $d/a_0=$
(a)0.8, (b)1.0, (c) 1.1, (d)1.2, (e)1.4, (f)1.7.
Only points with $|S(\vec{q})| > {1 \over 2} |S(\vec{q})|_{max}$
are plotted.  For large setback distances, a six-fold
symmetry appears, indicating the presence of quasi-long range
orientational order.}
\label{fig:sfactor}
\end{figure}

%%%FIG.9
\begin{figure}
\caption{Simulation of electric current for 3200 particles
with electric field stepped up every 4000 timesteps.
Magnitude of fluctuations is clearly much greater than
before electric field is turned on, indicating the
need for very long equilibration times between electric
field steps.}
\label{timecurrent}
\end{figure}

%%%FIG.10
\begin{figure}
\caption{Example of the pinning force vs. center of mass
shift $\Delta X$.  Sharp peaks measure the depinning electric
field, and give good agreeement with direct simulations of
electric field depinning.}
\label{pinf}
\end{figure}


\begin{references}
\bibitem{c1} The Burger's vector of a dislocation measures the amount
by which a circuit of steps between lattice sites
around the core of a dislocation fails
to close, if the steps in the circuit are chosen such that the
path would close for a lattice without a defect.
See Ref.~[22].
\bibitem{c3} The simplest assumption that generates this distribution
is that the disorder generates a Gaussian white noise distribution
of dislocations.  A more realistic assumption, that seems to
correspond quite well to our dislocation complexions in the
weak disorder limit, is that a given lattice site may have a
dislocation with probability $p$, or is a
defect-free lattice site with probability $1-p$.
The latter dislocation distribution gives results
identical to the former.
\bibitem{c4} In our simulations, we fix the temperature of the
electrons at some value, which is equivalent to assuming perfect
contact between the electron gas and a thermal reservoir.
Electron heating thus does not occur in our simulations.
\bibitem{c5} Occasionally,  electrons in the absence
of quenched disorder may freeze into
a crystal with an orientation that does not
match our boundary conditions.  In this situation, a small
number of defects will be incorporated in the ground state.
The number of such defects is always much smaller than the
number that we find in the presence of impurities, and we
believe that such possible global misorientations are not
relevant for the levels of disorder we have investigated.
\bibitem{c7} Since our error bars are relatively large
for larger values of $d$, and the scale in Fig.~\ref{fig:efield}
is logarithmic,
such an extrapolation is subject to very large uncertainties.
\end{references}
\end{document}